%%%%%%%If you do not have the msbm fonts, delete the following 4 lines
%\font\mybb=msbm10 at 12pt
%\def\bb#1{\hbox{\mybb#1}}
%%\def\Z {\bb{Z}}

%%%%%%%%%%%%
%%%and replace with the following 2 lines (without %)
\def\Z {Z}

%%%%%%%%%%

\def \aa {\alpha}
\def \bb {\beta}

\def \dd {\delta}

\def \ll {\lambda}

\def \rr {\rho}

\def \th {\theta}

\def \fff {\Phi}

\def \uuu {\Upsilon}

\def\bbb{{\bar \beta}}
\def\aab{{\bar \alpha}}

\def\ab{{\bar \alpha}}
\def\gb{{\bar \gamma}}

\def \2 {{1 \over 2}}
\def \3 {{1 \over 3}}
\def \4 {{1 \over 4}}
\def \5 {{1 \over 5}}
\def \6 {{1 \over 6}}
\def \7 {{1 \over 7}}
\def \8 {{1 \over 8}}
\def \9 {{1 \over 9}}
\def \0 { \infty}

\def\++ {{(+)}}
\def \- {{(-)}}
\def\+-{{(\pm)}}

\def\ek {\eqn\abc$$}

\def \pa {\partial}

  %%%%%%%This requires the PHYZZX.TEX macropackage

\tolerance=10000
\input phyzzx
%%%%%%%%%%%%%%%%%%%%%%%%%%%%%%%%%%%%%%%%%%%%%%%%%%%%%%%%%%%%%%%%%%%%

 \def\unit{\hbox to 3.3pt{\hskip1.3pt \vrule height 7pt width .4pt \hskip.7pt
\vrule height 7.85pt width .4pt \kern-2.4pt
\hrulefill \kern-3pt
\raise 4pt\hbox{\char'40}}}

\def\gij{g_{ij}}
\def\bij{b_{ij}}

\def\nup#1({Nucl.\ Phys.\  {\bf B#1}\ (}

\def\tr{{\rm tr}}

%%%%%%%%%%%%%%%%%%%%%%%%%%%%%%%%%%%%%%%%%%%%%%%%%%%%%%%%%%%%%%%%%%%%
%%%%%%%%%%%%%%%%%%%%%%%%%%%%%%%%%%%%%%%%%%%%%%%%%%%%%%%%%%%%%%%%%%%%

\REF\witten{E. Witten, ``Some Comments on String Dynamics'',
hep-th/9507121, Contributed to STRINGS 95:
Future Perspectives in String Theory, Los Angeles, CA, 13-18 Mar
1995.}
\REF\gstring{N. Seiberg, ``New Theories in Six Dimensions and
Matrix Description of M-theory on $T^5$ and $T^5/\Z_2$'',
hep-th/9705221.}%
\REF\seibhiggs{O. Aharony, M. Berkooz, S. Kachru, N. Seiberg, and
E. Silverstein, ``Matrix description of interacting theories in six
dimensions,'' hep-th/9707079}
\REF\stromopen{A. Strominger, ``Open p-branes,'' hep-th/9512059 }
\REF\town{P.K. Townsend, Phys. Lett. {\bf B373} (1996) 68}.
\REF\MK{D. Kutasov and E. Martinec, Nucl. Phys. {\bf B477} (1996) 652;
hep-th/9602049.}
\REF\MKO{D. Kutasov and E. Martinec and M. O'Loughlin, Nucl. Phys. {\bf B477}
(1996) 675;
hep-th/9603116.}
\REF\Mucsb{E. Martinec,  hep-th/9608017.}
\REF\MKA{D. Kutasov and E. Martinec,  hep-th/9612102.}
\REF\hull{C.M. Hull, hep-th 9702067.}
\REF\Mmat{E. Martinec,  hep-th/9706194.}
\REF\HW{C.M. Hull and E. Witten, Phys. Lett. {\bf 160B} (1985)  398.}
\REF\Hsig{ C.M. Hull, Nucl. Phys. {\bf B267} (1986)  266.}
\REF\Hsiga{ C.M. Hull,  Phys. Lett. {\bf 178B} (1986) 357.}
\REF\Hsigb{ C.M. Hull, in the Proceedings of the First Torino Meeting on
Superunification and Extra Dimensions,  edited by R. D'Auria and P. Fr\' e,
(World Scientific, Singapore, 1986).}
\REF\Van{ C.M. Hull, in Super Field Theories  (Plenum, New York, 1988), edited
by H. Lee and G. Kunstatter.}
%\REF\Mmat{T. Banks, W. Fischler, S. Shenker and L. Susskind, hep-th/9610043.}
\REF\tset{A. Tseytlin, hep-th/9701125.}
\REF\Bch{R. Bott and S-S. Chern, Acta Math {\bf 114} (1965) 71.}
\REF\Bcha{R. Bott and S-S. Chern, Essays on Topology and Related Topics,
Springer-Verlag (1970)
48.}
\REF\Don{S. Donaldson, Proc. Lond. Math. Soc., {\bf 50} (1985) 1; V. Nair and
J. Shiff, Nucl. Phys. {\bf B371} (1992) 329; Phys.
Lett. {\bf 246B} (1990) 423.}
 \REF\HP{P.S. Howe and G. Papadopoulos, Nucl. Phys. {\bf B289} (1987) 264.}
\REF\mohab{M. Abou Zeid and C.M. Hull, hep-th/9612208.}

%%%%%%%%%%%%%%%%%%%%%%%%%%%%%%%%%%%%%%%%%%%%%%%%%%%%%%%%%%%%%%%%%%%%

\Pubnum{ \vbox{  \hbox {QMW-97-26}  \hbox{hep-th/9708048}} }
\pubtype{}
\date{August, 1997}

\titlepage

\title {\bf  Non-Abelian Gravity and  Antisymmetric Tensor Gauge Theory}

\author{C.M. Hull}
\address{Physics Department,
Queen Mary and Westfield College,
\break
Mile End Road, London E1 4NS, U.K.}

\vskip 0.5cm

\abstract {
A non-abelian generalisation of a theory of gravity coupled to a 2-form gauge
field and a dilaton is found, in which the metric and 3-form field strength are
Lie algebra-valued.
In the abelian limit, the curvature with torsion is self-dual in four
dimensions, or  has $SU(n)$ holonomy in $2n$ dimensions. The coupling to
self-dual Yang-Mills fields in 4 dimensions, or their higher dimensional
generalisation, is discussed. The abelian theory is the effective action for
(2,1) strings,  and the non-abelian generalisation is relevant to the study of
coincident branes in the (2,1) string approach to M-theory.  The theory is
local when expressed in terms of a vector pre-potential.}

\endpage

\chapter {Introduction}

  Maxwell theory has a straightforward non-abelian extension to
Yang-Mills theory, but it has so far proved impossible to generalise this to
obtain non-abelian
anti-symmetric gauge theories. The abelian case involves an n-form gauge field
$b$ with abelian transformation
 $\dd b = d \aa$ and field strength $H=db$, but it seems impossible to find a
non-abelian generalisation that is local in
terms of $b$ unless $n=1$. However, it is now believed that there should be a
supersymmetric theory in $5+1$ dimensions
that includes a self-dual 2-form gauge field $b$, with $H=*H$ (plus corrections
involving scalar fields)
in the abelian limit, but
  which has a
non-abelian  generalisation that dimensionally reduces to Yang-Mills theory in
$4+1$ dimensions [\witten-\town].
Such theories arise from the world-volume theory of the M-theory 5-brane, which
becomes non-abelian in  the limit of coincident
5-branes [\stromopen,\town],  or from the low-energy-limit of the type IIB
string compactified on $K3$,
which has enhanced non-abelian gauge symmetry when two-cycles of the $K3$
degenerate [\witten].
It has been argued that this theory should exist as a consistent quantum theory
[\gstring].
Although we will not have anything to say about this case here, we will be able
to find a
non-abelian version of a related theory, which may have implications for the
six-dimensional tensor theory.

Martinec and Kutasov [\MK-\MKA] have argued that the superstring with (2,1)
world-sheet supersymmetry gives rise to the various branes and vacua of
M-theory and string theory.
The
(2,1) string is a theory of gravity plus an anti-symmetric tensor gauge field
and a Yang-Mills field in 2+2 dimensions, with
a generalised self-duality condition on both the curvature with torsion and the
Yang-Mills field strength.
The field equations were obtained in [\Hsig] and the action was given in
[\MKA,\hull], and a similar action can be used to
describe fields in
$10+2$ dimensions, before null reduction. This gives rise to the various branes
[\MK-\MKA];
for example reducing to a $1+1$ dimensional   real subspace of the $2+2$
dimensional space gives a Born-Infeld type string
action, which can be associated with that of a D-string [\Mucsb,\MKA].
 As coincident branes have non-abelian gauge symmetries, it should be
the case that the (2,1) string action should also have a non-abelian
generalisation, which would involve a non-abelian
generalisation of gravity plus an anti-symmetric tensor gauge field.
We will show that this is indeed the case,
giving a theory in which the metric and 2-form gauge field take values in a Lie
group. This may be of import for
the geometry of matrix theory, especially in view of the proposed relation
between matrix theory and the (2,1) string [\Mmat].

In (2,1) geometry, the metric and antisymmetric tensor are given in terms of a
vector $B$, with
$$  g_{\alpha \bar \beta} =i( \partial _ \alpha \bar B_ {\bar \beta} -  \bar
\partial _
{\bar \beta}
 B_ \alpha)
\eqn\aba$$
in complex coordinates. This has the obvious gauge symmetry $\dd B=d \aa$. It
has a natural non-abelian
generalisation in which $B$ becomes a non-abelian gauge field with
$\dd B=d \aa+[B,\aa]$ and field strength ${\cal F}
=dB+B^2$, so that \aba\ is replaced by $$g_{\alpha \bar \beta} =i{\cal
F}_{\alpha \bar \beta}.
\eqn\aba$$
The 2-form gauge field is also given in terms of $B$ and has a non-abelian
generalisation.
A gauge covariant theory can then be written in terms of the prepotential $B$,
which can be used to define a gauge-covariant
derivative etc. In particular, the real reduction of the $2+2$ dimensional
theory gives a non-abelian action of the D-string
type.  The resulting theory is local in terms of  the pre-potential $B$.

The Lie-algebra-valued metric can be used to define a line-element for a
manifold whose coordinates are G-valued matrices;
this may have applications to matrix theory.
However, the theories described here are   non-abelian generalisations of the
metric and 2-form gauge field in a standard
space-time with commuting coordinates. They are invariant under the standard
general coordinate transformations.

\chapter{The (2,1) Supersymmetric Sigma-Model}

The (1,1) supersymmetric  sigma-model  defined in a background  with metric
$\gij$, anti-symmetric tensor $\bij$
and dilaton $\Phi$
is conformally invariant at one-loop if the background fields satisfy
$$
R^{(+)} _{ij}- \nabla_{(i} \nabla _{j)} \Phi- H_{ij}{}^k \nabla _k\Phi
=0\eqn\conf$$
where $
R^{(+)} _{ij}$ is the Ricci tensor for a connection with torsion.
We define the connections with torsion
$$ \Gamma ^{(\pm) i}_{jk}=
\left\{ {i \atop jk} \right\}
\pm H^i_{jk}\eqn\con$$
where  $\left\{ {i \atop jk} \right\}$ is the Christoffel connection and the
torsion tensor is
$$
H_{ijk}= {3 \over 2}\partial_{[i}b_{jk]}.
\eqn\abc$$
The
curvature and Ricci tensors with torsion are
$$ R^{(\pm) k} {}_{lij}= \partial_i \Gamma^{(\pm) k}_{jl}-\partial_j
\Gamma^{(\pm) k}_{il}+ \Gamma ^{(\pm) k}_{im} \Gamma ^{(\pm) m} _{jl}-
 \Gamma ^{(\pm) k}_{jm} \Gamma ^{(\pm) m} _{il}, \qquad R^{(\pm)}  _{ij}=
R^{{(\pm)}
k}{}_{ikj}\eqn\cur$$ The equation \conf\ can be obtained from varying the
action
$$S= \int d^D x \, e^{-2\Phi} \sqrt{|g|}\left( R- {1\over 3} H^2+ 4 (\nabla
\Phi)^2\right).
\eqn\acttar$$

The sigma model is invariant under (2,1) supersymmetry [\HW-\Van] if the target
space is even
dimensional ($D=2n$) with a complex
structure
$J^i{}_j$    which is covariantly constant
$$ \nabla ^{(+)}_k J^i{}_j=0\eqn\covj$$
with respect to the connection $\Gamma^{(+)}$ defined in \con, and with respect
to which the metric is hermitian, so that
$J_{ij}\equiv g_{ik}J^k{}_j$ is antisymmetric.

It is useful to introduce complex coordinates $z^\alpha, \bar z^{\bar \beta}$
in which the line element is
$ds^2=2 g_{\alpha \bar \beta}d z^\alpha d \bar z^{\bar \beta}$ and the exterior
derivative decomposes
as $d=\partial +\bar \partial$.
 The   conditions for (2,1) supersymmetry imply that
the fundamental 2-form
 $$J={1 \over 2}J_{ij} d \phi ^i _ \Lambda d \phi ^j
=ig_{\alpha \bar \beta} dz^\alpha_\Lambda d \bar z^ {\bar \beta}\eqn\jis$$
satisfies
$$ i\partial \bar \partial J=0.
\eqn\jcon$$
Then
$$H=i(\partial - \bar \partial) J\eqn\hiss$$
{\it defines} a 3-form $H$ for which
the (0,3) and (3,0) parts
 vanish, and which is closed, $dH=0$, so that locally there is a 2-form $b$
with $H=db$.
Then
$$
H_{\aa \bb \gb}= \pa_{[\aa} g_{\bb] \gb}
\eqn\hist$$
Furthermore, \jcon\
implies  that locally there is a (1,0) form $k=k_\alpha d z^\alpha$ such that
$$J= i (\partial \bar k + \bar \partial k)\eqn\jisk$$
The metric and torsion potential are then given, in a suitable gauge,  by
$$
\eqalign{ g_{\alpha \bar \beta}&= \partial _ \alpha \bar k_ {\bar \beta} +
\bar \partial _
{\bar \beta}
 k_ \alpha
\cr
 b_{\alpha \bar \beta}&= \partial _ \alpha \bar k_ {\bar \beta} -  \bar
\partial _  {\bar \beta}
 k_ \alpha
\cr}
\eqn\kgeom$$
Setting $k_\aa=iB_\aa$ recovers \aba.
If $k_\alpha = \partial_\alpha K$ for some $K$, then the torsion vanishes and
the manifold is Kahler with Kahler potential $K$,
but if  $dk \ne 0$ then the space is a hermitian manifold of the type
introduced in [\HW].  The metric and torsion are invariant
under [\mohab]
$$ \delta k_\alpha = i \partial_ \alpha \chi + \theta_\alpha\eqn\ksym$$
where $\chi$ is real and $\theta_\alpha$ is holomorphic, $\partial_{\bar \beta}
\theta_ \alpha=0$.
Under these transformations, $g_{\aa \bbb} $ is invariant, while $b_{\aa\bbb}$
as defined in \kgeom\
transforms as
$$\dd b_{\aa \bbb} = -2i \pa _\aa \bar \pa _\bbb \chi
\ek
which is an antisymmetric gauge transformation
$$\dd b _{ij} = \pa _{[i} \ll_{j]}
\eqn\btrans$$
with parameter $\ll_\aa =2i \pa _\aa \chi$.
The equation \hiss\ only defines $b$ up to a gauge transformation $b \to
b+d\ll$. Such a gauge transformation can be used
to chose $b$ to be a (1,1) form as in \kgeom, or a
$(2,0)$ form plus a $(0,2)$ form given by
$$b_{\aa\bb}=\pa _{[\aa} k_{\bb]}
, \qquad
\bar b_{\aab\bbb}=\pa _{[\aab} \bar k_{\bbb]}
\eqn\bgeom$$
with $H_{\aa \bb \gb} $ given by
$$H_{\aa \bb \gb} =\pa _{\gb} b_{\aa \bb}
\eqn\hisb$$
Under \ksym,
$$ \dd b_{\aa\bb}= \pa_{[\aa} \theta_{\bb ]}
\ek
which is an antisymmetric gauge transformation \btrans\ with parameter
$\ll_\aa = \theta _\aa$.

It will be useful to define the vector
$$v^i=H_{jkl}J^{ij}J^{kl}
\eqn\vis$$
together with  the $U(1)$ part of the curvature
$$C^{(+)}_{ij}=J^l{}_kR^{{(+)} k} {}_{lij}\eqn\cis$$
and the $U(1)$ part of the connection \con\
$$\Gamma^{(+)} _i=J^k {}_j\Gamma^{(+)j}_{ik}=i( \Gamma^{(+)\alpha}_{i \alpha}-
\Gamma^{+\bar \alpha}_{i\bar \alpha})\eqn\uip$$
In a complex coordinate system, \cis\ can be written as
$C^{(+)} _{ij}= \partial _i \Gamma^{(+)} _j - \partial_j \Gamma^{(+)} _i$.

If the metric has Euclidean signature, then the holonomy of any metric
connection (including $\Gamma^{(\pm)}$)
is contained in $O(2n)$, while if it has signature $(2n_1,  2n_2)$ where
$n_1+n_2=n$, it will be in
$O(2n_1,2n_2)$. The holonomy ${\cal H}(\Gamma^{(+)})$ of the connection
$\Gamma^{(+)}$ is contained
in $U(n_1,n_2)$. It will be contained in $SU(n_1,n_2)$ if in addition
$$C^{(+)} _{ij}=0\eqn\ciso$$
where the $U(1)$ part of the curvature is given by \cis.
As $C_{ij}$ is a representative of the first Chern class, a necessary condition
for
this is the
vanishing of the first Chern class.

It was shown in [\Hsig] that geometries for which
$$\Gamma^{(+)} _i=0\eqn\fieldeq$$
in some suitable choice of coordinate system  will satisfy
the one-loop conditions \conf, provided the dilaton is chosen as
$$ \Phi= - \2 \log | det g_{\alpha \bar \beta}|
\eqn\fiis$$
which implies
$$\partial _i \fff= v_i
\eqn\fisv$$
 Moreover, the one-loop dilaton
field equation is also satisfied for compact manifolds, or for non-compact ones
in which $\nabla
\Phi$ falls off sufficiently fast [\hull]. This implies that ${\cal
H}(\Gamma^{(+)})
\subseteq SU(n_1,n_2)$ and these geometries generalise the Kahler Ricci-flat or
Calabi-Yau
geometries, and reduce to these in the special case in which
$H=0$.
These are not the most general solutions of \conf\ [\hull].

The equation \fieldeq\ can be viewed as a field equation for the potential
$k_\alpha$.
It can be obtained by varying the action [\MKA,\Mucsb,\hull]
$$
S=\int d^{D} x \sqrt{ | det g_{\alpha \bar \beta}}|
\eqn\act$$
where $g_{\alpha \bar \beta}$ is given in terms of $k_\alpha$ by \kgeom.
It can be rewritten as
$$
S=\int d^{D} x |   det g_{ij}|^{1/4}
\eqn\acta$$
which is non-covariant, as the field equation \fieldeq\ was obtained in a
particular coordinate system.
However, it is invariant under volume-preserving diffeomorphisms.

\chapter{Coupling to Yang-Mills Fields}

This theory of gravity can be coupled to (conventional) Yang-Mills fields
[\MKA,\hull].
 If $A$ is a connection on a holomorphic
vector bundle with structure group $H$
over such a hermitian geometry,
then we can define the Bott-Chern form $\Upsilon $ [\Bch] (constructed  in
[\Bcha,\Don,\HP]) by
$$\tr (F^2)= i \partial \bar \partial \Upsilon
\eqn\bcf$$
If   $H$ is abelian,
then $F^m=dA^m$ and there are real scalars $\phi ^m, \theta^m$ ($m=1,\dots,
rank(H)$) such that
$$\eqalign{
A^m &= d \th ^m + i(\pa -\bar \pa ) \phi ^m
\cr}
\eqn
\abd$$
and
the Bott-Chern form can be chosen to be
$$\Upsilon=-4i \partial   \phi^m  \bar \partial \phi^m$$

This can be used to define a sigma-model with (2,1) supersymmetry.
The 3-form field strength receives a Chern-Simons correction
$$
H = {1 \over 2}db+ \Omega (A)
\eqn\hcsp$$
where $d\Omega= trF^2$. Now
$$i \partial\bar \partial J= \tr (F)^2 \eqn\jeqp$$
and there is  (1,0) form $k$ such that
$$J=  \Upsilon  + i (\partial \bar k + \bar \partial k)\eqn\abc$$
and the metric and torsion potential  are given by
$$\eqalign{
 g_{\alpha \bar \beta} &= i  \Upsilon _{\alpha \bar \beta}+\partial _ \alpha
\bar k_
{\bar \beta} +
\bar \partial _  {\bar
\beta}
 k_ \alpha
\cr
 b_{\alpha \bar \beta} &= i  \chi _{\alpha \bar \beta} +\partial _ \alpha \bar
k_ {\bar
\beta} -
\bar \partial _  {\bar
\beta}
 k_ \alpha
\cr}
\eqn\erter$$
where $\chi$ is defined by
$$\Omega(A)= i(\pa - \bar \pa ) \Upsilon + d \chi
\eqn\csbc$$
The field equations can be obtained by varying the action \acta.
The Yang-Mills   equation is
$$ J^{ij}F_{ij}=0\eqn\uyau$$
This can be generalised to the case of (2,0) geometry,
but requires the introduction of gravitational Chern-Simons and Bott-Chern
terms [\hull].

It is sometimes useful to write the metric in terms of a fixed background
metric $\hat g_{\aa\bbb}$ (e.g. a flat metric)
which is given in terms of a potential $\hat k$ by
$\hat g_{\aa \bbb}= \pa _\aa \hat  k_ \bbb +\pa _\bbb \hat  k_\aa$, and a
fluctuation given in terms of  a vector field $B_i$ defined by
$$ B_\aa = -i (k_\aa-\hat k_\aa), \qquad
B_\ab = i(\bar k_\ab-\hat   k_\ab^*)
\eqn\abc $$
with field strength ${\cal F} =dB$. Then
$$   g_{\aa\bbb}=
\hat g_{\aa\bbb} +i{\cal F}_{\aa\bbb} +\uuu _{\aa\bbb}
\eqn\abc $$
The gauge
  symmetry \ksym\ has become  the usual gauge transformation of an abelian
gauge field
$$\dd B_i = \pa _ i \chi
\eqn\tihrue$$
and the action \act\
becomes
$$
S=\int d^{D} x
\sqrt{ | det ( \hat g_{\aa\bbb} +i{\cal F}_{\aa\bbb} +\uuu _{\aa\bbb} )|}
\eqn\actbf$$
which is similar to a
  Born-Infeld action. Note that
the (2,0) part of ${\cal F}$ is non-zero.

 \chapter{Non-Abelian Geometry}

A Born-Infeld action for an abelian gauge field
$$
S=\int d^{D} x
\sqrt{ | det ( G_{ij} +{ F}_{ij}   )|}
\eqn\actbfb$$
can be generalised to the non-abelian case in which $F$ takes values in some
Lie algebra
to give the action
$$
S=\int d^{D} x
Str \sqrt{ | det ( G_{ij} \unit  +{ F}_{ij}   )|}
\eqn\actbfa$$
where $Str$ denotes the symmetrised trace.  For D-brane actions, it was argued
in [\tset] that
this is the appropriate prescription, although in what follows it could be
replaced by any other
  suitable trace prescription.

The similarity between \actbfb\ and \actbf\ suggests that \actbf\ should also
have a non-abelian generalisation in which
$B_\aa$ becomes a non-abelian gauge field. The vector potential $k_\aa$ will
now be supposed to take values in some Lie
algebra $G$, so that $k_\aa=k_\aa^at_a$, where $t_a$ are Lie algebra
generators.
The abelian transformation \ksym\
$$ \delta k_\alpha = i \partial_ \alpha \chi  \eqn\ksyma$$
will now be generalised to the non-abelian transformation
$$ \delta k_\alpha = i \partial_ \alpha \chi + [k_\aa , \chi] \eqn\ksymn$$
where $\chi=\chi^at_a$.
A gauge-covariant G-valued generalisation of the metric, $g_{\alpha \bar
\beta}=g_{\alpha \bar \beta}^at_a$, is given by
$$g_{\alpha \bar \beta}= \partial _ \alpha \bar k_ {\bar \beta} +  \bar
\partial _
{\bar \beta}
 k_ \alpha -i [k_\aa , \bar k_{\bar \bb}]
\eqn\gnb$$
The natural generalisation of the action \act\ is then
$$
S=\int d^{D} x  \, Str \, \sqrt{ | det g_{\alpha \bar \beta}}|
\eqn\actn$$
Defining
$$ B_\aa = -i k_\aa, \qquad
B_\ab = i\bar k_\ab
\eqn\abc $$
gives a potential in which \ksyma\ becomes the usual gauge transformation
$$\dd B_i = \pa _ i \chi +[B_i,\chi]
\eqn\tihruen$$
with field strength ${\cal F} =dB+{1\over 2}[B,B]$. Then the non-abelian metric
is given by the
(1,1) part of the field strength
$$   g_{\aa\bbb}=-
i{\cal F}_{\aa\bbb}
\eqn\abc $$

This can be generalised to include the coupling to Yang-Mills fields $A$ taking
values in a group $H$, which are taken as
singlets of $G$. Then
$$   g_{\aa\bbb}=\uuu _{\aa\bbb} \unit +
 i{\cal F}_{\aa\bbb}
\eqn\abc $$
where $\unit$ is the identity of $G$
and
the action \actn\
becomes
$$
S=\int d^{D} x  \, Str \,
\sqrt{ | det ( \uuu_{\aa\bbb} \unit +i{\cal F}_{\aa\bbb}   )|}
\eqn\actbf$$

Let $D_i$ be the covariant derivative constructed using the connection
$B_i=J_i{}^jk_j$, so that
$$
{\cal F}_{ij}=[D_i, D_j]
\ek
Then the expression \hist\ for $H$ in the abelian case can be generalised to a
gauge-covariant field strength
$$
H_{\aa  \gb \bb}= D_{[\aa} g_{\bb] \gb}=\pa_{[\aa} g_{\bb] \gb} +
[B_{[\aa}, g_{\bb] \gb}]
\ek
Using   the Bianchi identity $D_{[\aa}{\cal F}_{\bb \gb]}=0$, this can be
rewritten as
$$H_{\aa \bb \gb}=iD_{[\aa} {\cal F}_{\bb] \gb}= iD_\gb {\cal F}_{\aa \bb}
\ek
This can be thought of as a covariantisation of \hisb, so that
the antisymmetric tensor gauge field is now given by
$$ b_{\aa \bb}=i {\cal F}_{\aa \bb}
\eqn\ertyyey$$
with $b_{\aa\bbb}=0$. Then both $g_{ij}$ and $b_{ij}$ are invariant under
\gnb\ or \tihruen. The $\theta $ transformation in \ksym,
$$\dd k_\aa =\theta _\aa
\ek
with $\theta$ holomorphic remains a symmytry if it is now covariantly
holomorphic,
$$D_\bbb \theta _\aa=0
\ek
Then $g_{\aa \bbb}$ and $H_{\aa\bb\gb}$ are invariant under this, while $b$
transforms as
$$ \dd b_{\aa \bb }
= D_{[\aa} \th _{\bb]}
\ek
This is the non-abelian generalisation of the antisymmetric tensor gauge
symmetry $\dd b= d \ll$.
The gauge-invariant field equation for $H$ now takes the form
$D^iH_{ijk}=0+\dots $.

Consider now the special case in which the original geometry is Kahler, so that
$$B_\aa =-i \pa _\aa K
\eqn\ertw$$
in the abelian case. This implies that $B_\aa$ is a holomorphic connection,
with ${\cal F}_{\aa\bb}=0$.
The natural non-abelian generalisation is to take $B$ to be a connection on a
holomorphic vector bundle with structure group
$G$, so that ${\cal F}_{\aa\bb}=0$ and there is   locally a complex
Lie-algebra-valued prepotential $V$ such that
$$ B_{\aa} =  V^{-1} \pa _ \aa V
\ek
A gauge transformation can be used to set the real part of $V$ to zero, so that
in the abelian limit $V=\exp (-iK) $
and \ertw\ is recovered. The prepotential     $V$ transforms as
$$ V \to \bar \rr V g
\ek
under a gauge transformation parameterised by
the  group element $g$ and under a
pregauge transformation with holomorphic group-valued parameter $\rr(z)$, which
reduces to the Kahler gauge transformation
$\dd K =f(z) + \bar f(\bar z)$ in the abelian limit with $\rr = \exp (2if)$.
In this case, $H$ is zero.

Acknowledgements

I would like to thank Mohab Abou Zeid  for useful discussions.

\refout

\bye